\documentstyle[procl,psfig]{article}
  \newcommand{\ccaption}[2]{
    \begin{center}
    \parbox{0.85\textwidth}{
      \caption[#1]{\small{\it{#2}}}
      }
    \end{center}
    }

\def\be{\begin{equation}}
\def\ee{\end{equation}}
\def\bea{\begin{eqnarray}}
\def\eea{\end{eqnarray}}

\begin{document}

\title{SPIN EFFECTS IN LEPTON-NUCLEON SCATTERING: A THEORETICAL OVERVIEW}

\author{G. RIDOLFI}

\address{INFN Sezione di Genova, Via Dodecaneso 33, I-16146 Genova (Italy)}

\maketitle\abstracts{I review recent theoretical results on polarized
deep inelastic lepton nucleon scattering. Some specific issues,
like $Q^2$ evolution of structure functions, the small $x$ behaviour,
and the determination of polarized parton densities, are discussed in
some detail.}

\section{Introduction}
The physics of polarized nucleons has received increasing interest in the
past few years, as we had the opportunity to see during the
Workshop. Experimental progress in this field has been impressive,
and prospect for the next future are even more interesting. A description
of the present status of the field on the experimental side is given
in another chapter of these Proceedings;~\cite{Magnon} here I will concentrate
on the theoretical aspects.

The most important tool for studying the structure of polarized nucleons
is polarized deep-inelastic scattering, and most of the work presented
in our Workshop was devoted to the study of polarized structure functions;
I conclude this introduction recalling a few basics concepts in
this field. In Section~\ref{scaling},
I will discuss the problem of scaling violation
in polarized deep inelastic scattering, in the context of perturbation theory.
In Section~\ref{smallx} I will review the theoretical status
of the small $x$ behaviour
of polarized structure functions, and in Section~\ref{distrib}
I will discuss the extraction
of polarized parton densities. Finally, in Section~\ref{concl} I
present my conclusions.

The antisymmetric part of the hadronic tensor, relevant for polarized
deep inelastic lepton-nucleon scattering, is usually decomposed as
\be
\label{wmunu}
W_{\mu\nu}^A(x,Q^2)=\frac{m}{p\cdot q}\epsilon_{\mu\nu\rho\sigma}q^\sigma
\left[s^\rho g_1(x,Q^2)+\left(s^\rho-\frac{q\cdot s}{p\cdot q}\,p^\rho\right)
g_2(x,Q^2)\right],
\ee
where $m,s,p$ are the nucleon mass, spin and four-momentum respectively,
$q$ is the virtual photon momentum, $Q^2=-q^2$ and
$x=Q^2/(2p\cdot q)$. This parametrization applies to the case of neutral
lepton currents; for charged current scattering three more structure functions
must be introduced.~\cite{AnselminoBluemlein}
It is easy to see that, for longitudinal polarizations
$s$, $W_{\mu\nu}^A$ is dominated by the $g_1$ term in the Bjorken limit
of large $Q^2$ at fixed $x$. 

The first moment of $g_1$,
\be
\Gamma_1(Q^2)=\int_0^1 dx \,g_1(x,Q^2) 
\ee
is a quantity of great interest, because it is related to the nucleon
matrix element of the quark axial current $j^\mu_5=\sum_{i=1}^{n_f}
\bar{\psi}_i \gamma^\mu\gamma_5 \psi_i$
whose experimental determination allows
important tests of theoretical predictions of perturbative QCD, and possibly
provides hints about non-perturbative effects.
There are two main difficulties in extracting $\Gamma_1$ from measurements
of $g_1$ at different values of $x$ and $Q^2$. First, all data point must 
be evolved to the same $Q^2$ before estimating the integral. Second, one
needs an extrapolation criterion to estimate the contribution to $\Gamma_1$
from unmeasured regions of the $x$ range. During the Workshop we have learned
how both difficulties can be overcome.

\section{Scale dependence}
\label{scaling}
In the parton model, improved by QCD radiative corrections, the structure 
function $g_1$ has the following expression:
\bea
\label{g1}
g_1(x,Q^2) & = & \frac{\langle e^2\rangle}{2}
        \int_x^1 \frac{dy}{y} \Bigl [ 
        C_q^{\rm S}(\frac{x}{y},\alpha_s(Q^2)) \Delta\Sigma(y,Q^2)
      \\
      & + & 2 n_f C_g(\frac{x}{y},\alpha_s(Q^2)) \Delta g(y,Q^2) 
  + C_q^{\rm NS}({\frac{x}{y}},\alpha_s(Q^2)) \Delta q^{\rm NS}(y,Q^2) \Bigr],
\nonumber
\eea
where $\langle e^2\rangle = \sum_{k=1}^{n_f} e_k^2/n_f$, and
$\Delta\Sigma$ and $\Delta q^{\rm NS}$ are the singlet and non-singlet
polarized quark distributions
\be
\Delta\Sigma(x,Q^2) =  \sum_{i=1}^{n_f} \Delta q_i(x,Q^2),
\qquad
\Delta q^{\rm NS}(x,Q^2)  =  \sum_{i=1}^{n_f} 
                ( e_i^2/\langle e^2\rangle -1)\Delta q_i(x,Q^2).
\label{qcd:qsns}
\ee
The coefficient functions $C_q^{\rm S,NS}$, $C_g$ have been computed
up to next-to-leading order~\cite{Kodaira} in QCD.
The polarized parton distributions are of non-perturbative nature, but
their scale dependence is determined in perturbation theory by polarized
Altarelli-Parisi equations.~\cite{AltarelliParisi}
A recent, important theoretical achievement in this context is the calculation
of the splitting functions at next-to-leading order in the polarized case,
performed independently by Mertig and van Neerven~\cite{Mertig} 
and by Vogelsang~\cite{Vogelsang}. The calculation of next-to-leading order
Altarelli-Parisi kernels is particularly difficult in the polarized case,
because the presence of the $\gamma_5$ matrix requires great care in 
the use of dimensional regularization. The existence of two independent 
calculations, performed with two different techniques (the operator
product expansion in ref.~\cite{Mertig} and the parton model in the light-like
axial gauge for ref.~\cite{Vogelsang}) puts this important result on a firm 
ground.

The problem of evolving $g_1$ data points to the same $Q^2$, which
has also been treated by methods based on the assumption
that the evolution of $g_1$ is approximately equal to that of
unpolarized structure functions~\cite{EllisKotikov}, can now
be consistently approached within next-to-leading order QCD.
This involves assigning parton distributions at an initial values of $Q^2$
in terms of a set of free parameters, using eq.~(\ref{g1}) and the 
evolution equations to compute $g_1$ for values of $x$ and $Q^2$
corresponding to data points, and performing a fit of the initial
condition parameters.
Such an analysis has been carried out by different
groups;~\cite{GS,BFR,GRSV} details on the results
have been presented during the Workshop, and can be found
in~\cite{Stirling,Forte}. In section~\ref{distrib} I will discuss
some specific aspects of these results.

\section{Small-$x$ behaviour}
\label{smallx}
The problem of small (and large) $x$ extrapolation of $g_1$ is closely
related to that of $Q^2$ evolution. In the approach of
refs.~\cite{GS,BFR,GRSV},
after the fit parameters have been determined, the $x$ behaviour
of $g_1$ outside the measured range is completely fixed, 
and the integral can be taken over the whole range from 0 to 1.
The associated uncertainty is controlled by the errors on the fit parameters
and implicitly by the functional form chosen for the parametrization. 
This procedure allows to estimate the contribution
to $\Gamma_1$ from the unmeasured region taking into proper account the effect
of $Q^2$ evolution.
Such effects are not taken into account if, for example, the
small-$x$ contribution to the first moment is estimated by assuming 
that in the small $x$ region $g_1$ is equal to the average of the last two
data points; in fact $Q^2$ evolution at a given value of $x=\bar{x}$ is
determined by convolution integrals, and therefore it
is affected by the value of the structure function at all values $x>\bar{x}$.

Evolution effects at small $x$ are particularly sensitive to higher
order perturbative corrections, which are enhanced by powers of
$\alpha_s \log^2 (1/x)$. It has been checked in ref.~\cite{BFR}
that this is not yet a problem in the $x$ range probed by present
experiments. 
However, the resummation of these contributions~\cite{Ermolaev},
although unsupported by appropriate factorization theorems, 
leads to a power-like rise of both the singlet and the non-singlet
components of $g_1$ as $x\to 0$. In the singlet case this rise
is so strong as to produce a divergent first moment of $g_1$.
The impact of these corrections on the evolution of
the next-to-leading order best-fit non-singlet parton 
distributions~\cite{GS,BFR,GRSV},
which already display a strong power-like rise at small $x$, is rather
small~\cite{BluemleinSX}.
However it is important to bear in mind that in the unpolarized case,
where factorization theorems do exist, and the double logs are known 
to cancel down to single logs, there still seems to be some disagreement 
between the results expected theoretically from resummations and the 
HERA data on $F_2$. 

\section{Polarized parton distributions}
\label{distrib}
As mentioned in Section~\ref{scaling}, in refs.~\cite{GS,BFR,GRSV}
a next-to-leading order fit to deep inelastic scattering data has been
performed; this procedure requires the determination of polarized parton
densities in the nucleon.
Measurements of $g_1$ are not sensitive to the individual quark distributions,
but only to their sum, weighted by their squared charges; for this reason,  
presently available data do not allow a satisfactory
determination of individual quark densities, and some {\it ad hoc} assumption
on the flavor structure has to be made; for example, in ref.~\cite{BFR}
all non-singlet quark distributions are assumed to have the same 
$x$ dependence, while in ref.~\cite{GS} an SU(3)-symmetric structure of
antiquark distributions is assumed.
Future data on semi-inclusive deep inelastic scattering~\cite{Anselmino}
may help to disentangle the distributions of individual flavours.

\begin{figure}[htb]
\begin{center}
\mbox{\psfig{file=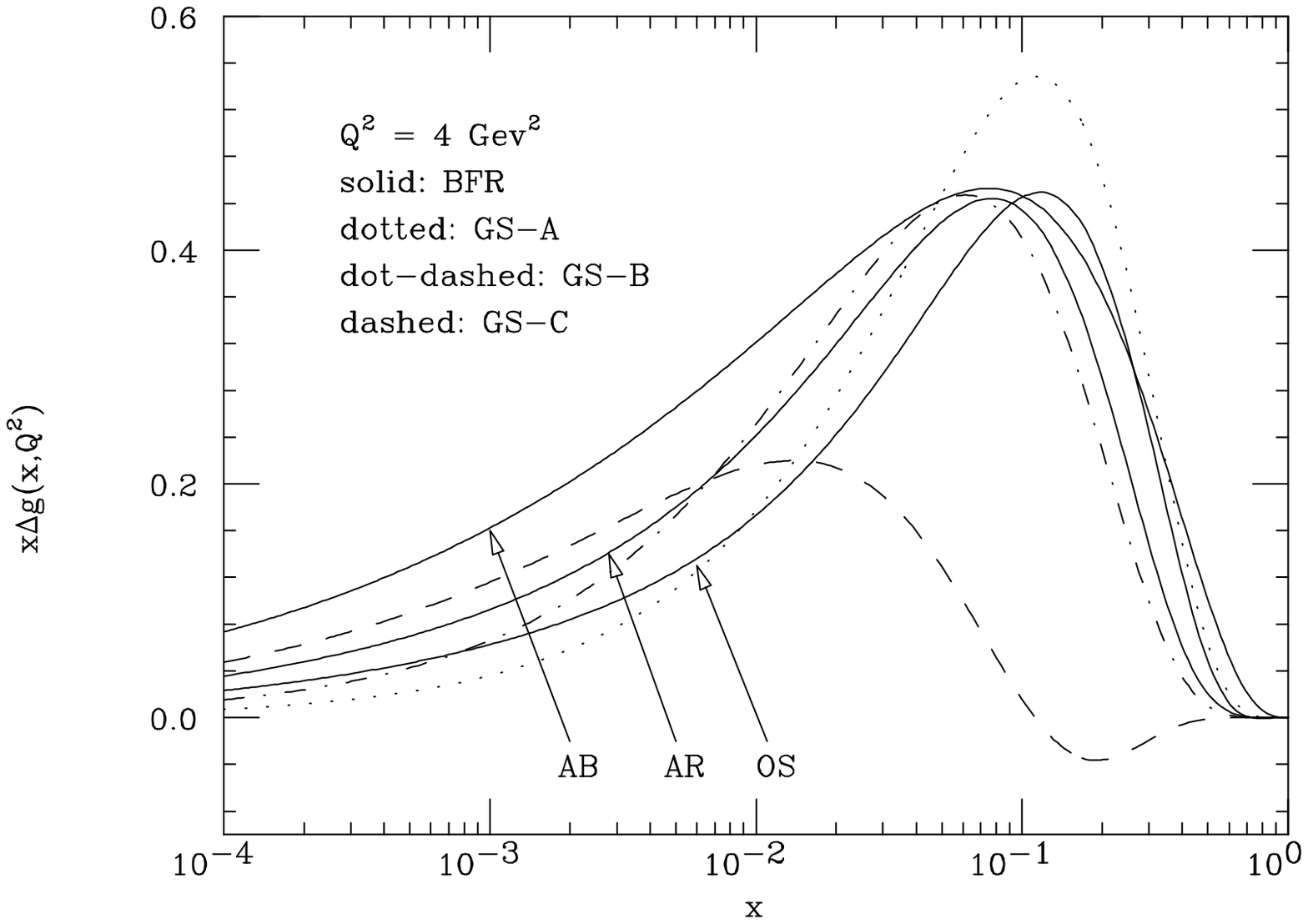,width=0.70\textwidth}}
\ccaption{}{\label{gluons}
Polarized gluon density in the proton in different parametrizations.
}
\end{center}
\end{figure}

The solution of the coupled Altarelli-Parisi equations 
in the singlet sector involves the determination of the polarized gluon
density $\Delta g$, which is therefore measured indirectly through
scaling violation, just as in the unpolarized case.
On the other hand, the polarized gluon distribution, and in particular the
value of its first moment, is particularly important here,
because it affects the determination of the singlet axial charge.
It is therefore interesting to know to what extent present data allow
a determination of $\Delta g$. In fig.~\ref{gluons} I present six
different parametrizations of $\Delta g$, obtained~\cite{GS,BFR}
with the procedure outlined above using
available data on polarized deep inelastic scattering with proton
and deuteron targets.

%%%%%%%%%%%%%%%%%%%%%%%%%%%%%%%%%%%%%%%%%%%%%%%%%%%%%%%%%%%%%%%%%%%%%
%%%%%%%%%%%%%%%%%%%%%%%%%%%%%%%%%%%%%%%%%%%%%%%%%%%%%%%%%%%%%%%%%%%%%
The three parametrizations of ref.~\cite{BFR}, labelled by AB, OS and AR,
differ because they were obtained in different subtraction 
(factorization) schemes
for collinear divergences; the curves labelled by GS-A, GS-B and GS-C,
instead, were all obtained in the $\overline{\rm MS}$ subtraction scheme,
and they differ because a different large$-x$ behaviour was chosen for
each of them. 

The subtraction scheme dependence is particularly subtle in the polarized case
because of the extra ambiguity related to the way chiral symmetry
is broken by the regularization procedure (see ref.~\cite{Forte}
for a complete discussion of this point). This is reflected in an
ambiguity in the size of the first moment of the gluon coefficient
function $C_g$ (which starts at order $\alpha_s$). 
In particular, the usual $\overline{\rm MS}$ prescription
in dimensional regularization must be supplemented
with suitable finite counterterms in order to recover non-singlet axial current
conservation. In this scheme, adopted in refs.~\cite{GS,GRSV},
the first moment of the gluon coefficient function $C_g^1$ vanishes.
In a different, widely-adopted class of schemes,
one fixes the finite counterterms in such a way that
the first moment of the polarized quark distribution
is scale independent, which implies $C_g^1=-{\alpha_s\over 4\pi}$.
In this class of schemes, the quark singlet distribution,
being conserved by evolution, can be interpreted in a natural way as the
total helicity carried by quarks.
The three factorization schemes used in ref.~\cite{BFR} belong to this class.
The quark distributions change by an amount proportional to $\Delta g$
when going from one class of schemes to the other; this
change is compensated by a corresponding modification of coefficient functions
and two-loop Altarelli-Parisi kernels, so that physical quantities like $g_1$
remain unaffected, up to next-to-next-to-leading order terms.

Because the 
evolution equations imply that at leading order the first moment of 
$\alpha_s \Delta g$ does not vanish at large $Q^2$,
this scheme dependence persists asymptotically and is potentially large
if the first moment of the gluon distribution is large.
On the other hand the gluon distribution,
and in particular its first moment, is not affected, at this order,
by the scheme change,
and therefore parametrizations of $\Delta g$ obtained
in different schemes, like those presented in fig.~\ref{gluons},
can be directly compared.
We see that five of the six parametrizations considered are quite close 
with each other in shape; the sixth one, namely the GS-C parametrization,
was obtained by forcing a particular large-$x$ behaviour of the initial 
conditions; the quality of the resulting fit is only slighlty worse than 
for the others. This suggests that present data are not precise enough to
determine the exact behaviour of $\Delta g$ in the whole $x$ range. However,
I would like to stress that enormous progress has been made
in the determination of $\Delta g$, and in particular that the
situation has significantly improved since
deuteron data have become available.
Furthermore, it is interesting to observe that, using presently published
data, the first moment of 
$\Delta g$ can be determined with an accuracy of about 50\%;
the result of ref.~\cite{BFR}, for example, is
\be
\label{etagbfr}
\eta_g\equiv\int_0^1 dx \Delta g(x,1\;{\rm GeV}^2)=1.52\pm0.74,
\ee
to be contrasted with the fact that proton data alone would even
be consistent~\cite{BFRa} with $\eta_g=0$. The fits of ref.~\cite{GS} are
consistent with the result in eq.~(\ref{etagbfr}).
The uncertainty quoted in eq.~(\ref{etagbfr}) will improve very much
in the near future, with the inclusion of recent deuteron data
from the SMC Collaboration, and particularly if polarized deep inelastic
scattering data will be collected at HERA. This possibility has been discussed
during our Workshop~\cite{Lichtenstadt}, and it has been concluded that 
an accuracy of about 30\% for $\eta_g$ 
is achieavable at HERA, assuming a
luminosity of 200~pb$^{-1}$ in the polarized configuration.
The important point here is that HERA measurements
will help in determining $\eta_g$ because they will allow measurements
of $g_1$ in the same $x$-range as previous experiments, but at much
larger values of $Q^2$.

It is apparent from the above discussion that a direct
experimental determination of $\Delta g$ would be of great interst,
because it would provide an independent
check on the results obtained from scaling violation.
There are several proposals in this direction.
A direct measurements of the polarized gluon distribution can be performed
by studying two-jet production with polarized beams at HERA, a process 
dominated by photon-gluon fusion~\cite{Mirkes} for relatively small $x$.
In ref.~\cite{Mirkes} the conclusion is reached that a good measurement 
of $\Delta g$ for $x$ below 0.2 is achievable at HERA 
in the polarized configuration, if the designed conditions
of energy and luminosity will be reached. This estimate is based on 
leading order calculations, since a next-to-leading order
calculation of the cross section for polarized dijet production
is lacking. 

A second possibility is the production of heavy quark pairs in
photon-proton collisions. This possibility is being consider by
the COMPASS collaboration in the deep-inelastic scattering 
regime~\cite{VonHarrach}.
It has been recently pointed out~\cite{FrixioneVogelsang}
that charm asymmetries can be measured at HERA in the polarized 
configuration, and that significant constraints on $\Delta g$ will
be put by this kind of measurements. Also in this case, a next-to-leading
order calculation is unfortunately not available.

\section{Conclusion and outlook}
\label{concl}
Important progress has been made in the study of the structure of
polarized nucleons in the recent past.
The structure function $g_1$ is the quantity which has undergone
the most detailed analyses, since it is relatively easily accessible
from the experimental point of view.
The perturbative study of $g_1$ has now reached a level of accuracy
which is comparable to the one we are used to in the case of unpolarized
structure functions,
since both coefficient functions and Altarelli-Parisi kernels are known 
at next-to-leading order. The results of the different analyses performed
are consistent with each other, and indicate that the density of polarized
gluons in the nucleons is significantly different from zero.
Future data will considerably improve our
knowledge of polarized parton distributions; scaling violation will be studied
at a higher level of accuracy, and direct measurements of polarized asymmetries
are expected in the near future. 

\section*{Acknowledgements} 
I wish to thank Stefano Forte for his considerable help
in summarizing the results presented during the Workshop,
and Richard Ball for the pleasant and
fruitful work done together in the past two years.
I would also like to thank the Organizing Committee of DIS '96 for the
kind and efficient hospitality in Rome.


\section*{References}
\begin{thebibliography}{99}
\bibitem{Magnon}
  A.~Magnon, these Proceedings.
\bibitem{AnselminoBluemlein}
  M~Anselmino, P.~Gambino and J.~Kalinowski, Z. Phys. {\bf C64} (1994) 267;\\
  J.~Bl\"umlein and N.~Kochelev, these Proceedings.
\bibitem{Kodaira}
  J.~Kodaira et al., Phys. Rev. {\bf D20} (1979) 627.
\bibitem{AltarelliParisi}
  G.~Altarelli ang G.~Parisi, Nucl. Phys. {\bf B126} (1977) 298.
\bibitem{Mertig}
  R.~Mertig and W.L.~van~Neerven, Z. Phys. {\bf C70} (1996) 637.
\bibitem{Vogelsang}
  W.~Vogelsang, preprint RAL-TR-96-020, hep-ph/9603366 (1996); these
  Proceedings.
\bibitem{EllisKotikov}
  J.~Ellis and M.~Karliner, Phys. Lett. {\bf B313} (1993) 131;\\
  A.V.~Kotikov and D.V.~Peshekhonov, these Proceedings.
\bibitem{GS}
  T.~Gehrmann and W.~J.~Stirling, Phys. Rev. {\bf D53} (1996) 6100. 
\bibitem{BFR}
  R. D. Ball, S. Forte and G. Ridolfi, Phys. Lett. {\bf B378} (1996) 255.
\bibitem{GRSV}
  M.Gl\"uck,  E. Reya, M. Stratmann and W. Vogelsang, Phys. Rev. {\bf D53} 
  (1996) 4775 .
\bibitem{Stirling}
  T.~Gehrmann and W.J.~Stirling, these Proceedings.
\bibitem{Forte}
  S.~Forte, R.D.~Ball and G.~Ridolfi, these Proceedings.
\bibitem{Ermolaev}
  J.~Bartels, B.I.~Ermolaev and M.G.~Ryskin, these Proceedings.
\bibitem{BluemleinSX}
  J.~Bl\"umlein, S.~Riemersma and A.~Vogt, these Proceedings.
\bibitem{Anselmino} 
  M.~Anselmino, M.~Boglione, J.~Hansson and F.~Murgia, these Proceedings.
\bibitem{BFRa}
  R. D. Ball, S. Forte and G. Ridolfi, Nucl. Phys. {\bf B444} (1995) 287.
\bibitem{Lichtenstadt}
  J.~Lichtenstadt et al., these Proceedings.
\bibitem{Mirkes}
  J.~Feltesse, F.~Kunne and E.~Mirkes, these Proceedings.
\bibitem{VonHarrach}
  D.~Von Harrach, these Proceedings.
\bibitem{FrixioneVogelsang}
  S.~Frixione and G.~Ridolfi, Phys. Lett. {\bf B383} (1996) 227;\\
  M.~Stratmann and W.~Vogelsang, DO-TH-96/10, RAL-TR-96-033,
  hep-ph/9605330.
\end{thebibliography}
\end{document}